\documentclass[aps,preprint,prl,showpacs]{revtex4}
\usepackage{graphicx}

\begin{document}

\def \reofour {$\mathrm{(TMTSF)_{2}ReO_4}$\,}
\def \clofour {$\mathrm{(TMTSF)_{2}ClO_4}$\,}
\def \areofour {$\mathrm{ReO_4^-}$\,}
\def \aclofour {$\mathrm{ClO_4^-}$\,}
\def \pfsix {$\mathrm{(TMTSF)_{2}PF_6}$\,}
\def \asfsix {$\mathrm{(TMTSF)_{2}AsF_{6}}$}
\def \tmtsfx {$\mathrm{(TMTSF)_{2}X}$\,}
\def\tc{$T_{c}$\,}
\def\pc{$P_{c}$\,}
\def\tmp{$\mathrm{(TMTSF)_{2}PF_{6}}$\,}
\def\t6as{$\mathrm{(TMTSF)_{2}AsF_{6}}$}
\def\tmx{$\mathrm{(TMTSF)_{2}ClO_{4(1-x)}ReO_{4x}}$}\,
\def\tmc{$\mathrm{(TMTSF)_{2}ClO_{4}}$\,}
\def\tms{$\mathrm{(TMTSF)_{2}AsF_{6(1-x)}SbF_{6x}}$}\,
\def\tmps{$\mathrm{(TMTTF)_{2}PF_{6}}$\,}
\def\tmttfsbf6{$\mathrm{(TMTTF)_{2}SbF_{6}}$\,}
\def\tmttfasf6{$\mathrm{(TMTTF)_{2}AsF_{6}}$\,}
\def\tmttfbf4{$\mathrm{(TMTTF)_{2}BF_{4}}$\,}
\def\tmtsfreo4{$\mathrm{(TMTSF)_{2}ReO_{4}}$\,}
\def\tq{$\mathrm{TTF-TCNQ}$\,}
\def\tsq{$\mathrm{TSeF-TCNQ}$}\,
\def\qnq{$(Qn)TCNQ_{2}$}\,
\def\R{$\mathrm{ReO_{4}^{-}}$}  
\def\C{$\mathrm{ClO_{4}^{-}}$}
\def\P{$\mathrm{PF_{6}^{-}}$}
\def\tqr{$\mathrm{TCNQ^\frac{\cdot}{}}$\,}
\def\nmpq{$\mathrm{NMP^{+}(TCNQ)^\frac{\cdot}{}}$\,}
\def\q{$\mathrm{TCNQ}$\,}
\def\nmp{$\mathrm{NMP^{+}}$\,}
\def\f{$\mathrm{TTF}\,$}
\def\tc{$T_{c}$\,}
\def\tc{$T_{c}$\,}
\def\nmq{$\mathrm{(NMP-TCNQ)}$\,}
\def\ts{$\mathrm{TSeF}$}
\def\tsm{$\mathrm{TMTSF}$\,}
\def\tst{$\mathrm{TMTTF}$\,}
\def\tmp6{$\mathrm{(TMTSF)_{2}PF_{6}}$\,}
\def\tms2x{$\mathrm{(TMTSF)_{2}X}$}
\def\as{$\mathrm{AsF_{6}}$}
\def\sb{$\mathrm{SbF_{6}}$}
\def\pf{$\mathrm{PF_{6}}$}
\def\re{$\mathrm{ReO_{4}}$}
\def\ta{$\mathrm{TaF_{6}}$}
\def\cl{$\mathrm{ClO_{4}}$}
\def\4fb{$\mathrm{BF_{4}}$}
\def\ttdm{$\mathrm{(TTDM-TTF)_{2}Au(mnt)_{2}}$}
\def\edt{$\mathrm{(EDT-TTF-CONMe_{2})_{2}AsF_{6}}$}
\def\tfx{$\mathrm{(TMTTF)_{2}X}$\,}
\def\tsx{$\mathrm{(TMTSF)_{2}X}$\,}
\def\ttftcnq{$\mathrm{TTF-TCNQ}$\,}
\def\ttf{$\mathrm{TTF}$\,}
\def\tcnq{$\mathrm{TCNQ}$\,}
\def\bedtttf{$\mathrm{BEDT-TTF}$\,}
\def\reo4{$\mathrm{ReO_{4}}$}
\def\bedtttfreo4{$\mathrm{(BEDT-TTF)_{2}ReO_{4}}$\,}
\def\et2i3{$\mathrm{(ET)_{2}I_{3}}$\,}
\def\et2x{$\mathrm{(ET)_{2}X}$\,}
\def\ket2x{$\mathrm{\kappa-(ET)_{2}X}$\,}
\def\cuncnbr{$\mathrm{Cu(N(CN)_{2})Br}$\,}
\def\ket2x{$\mathrm{\kappa-(ET)_{2}X}$\,}
\def\cuncncl{$\mathrm{Cu(N(CN)_{2})Cl}$\,}
\def\cuncs{$\mathrm{Cu(NCS)_{2}}$\,}
\def\betsfecl4{$\mathrm{(BETS)_{2}FeCl_{4}}$\,}
\def\bets{$\mathrm{BETS}$\,}
\def\hc2{$H_{c2}(T)$}
\def\et{$\mathrm{ET}$\,}
\def\tmm{$\mathrm{TM}$\,}
\def\tmtsf{$\mathrm{(TMTSF)}$\,}
\def\tmttf{$\mathrm{(TMTTF)}$\,}
\def\tm2x{$\mathrm{(TM)_{2}X}$\,}
\def\t1{${1/T_1}$\,}

\title{The metallic transport of \tms2x organic conductors close to the superconducting phase}
\author{P. Auban-Senzier}
\email{senzier@lps.u-psud.fr}
\affiliation{Laboratoire de Physique des Solides, UMR 8502 CNRS Universit\'e Paris-Sud, 91405 Orsay, France}

\author{D. J\'erome}
\email{jerome@lps.u-psud.fr}
\affiliation{Laboratoire de Physique des Solides, UMR 8502 CNRS Universit\'e Paris-Sud, 91405 Orsay, France}

\author{N.~Doiron-Leyraud}
\email{ndl@physique.usherbrooke.ca}
\affiliation{D\'epartement de Physique and RQMP, Universit\'e de Sherbrooke, Sherbrooke, Qu\'ebec, J1K 2R1, Canada}

\author{S. Ren\'e de Cotret}
\affiliation{D\'epartement de Physique and RQMP, Universit\'e de Sherbrooke, Sherbrooke, Qu\'ebec, J1K 2R1, Canada}

\author{A.  Sedeki}
\affiliation{D\'epartement de Physique and RQMP, Universit\'e de Sherbrooke, Sherbrooke, Qu\'ebec, J1K 2R1, Canada}

\author{C.~Bourbonnais}
\email{cbourbon@physique.usherbrooke.ca}
\affiliation{D\'epartement de Physique and RQMP, Universit\'e de Sherbrooke, Sherbrooke, Qu\'ebec, J1K 2R1, Canada}

\author{L.~Taillefer}
\email{ltaillef@physique.usherbrooke.ca}
\affiliation{D\'epartement de Physique and RQMP, Universit\'e de Sherbrooke, Sherbrooke, Qu\'ebec, J1K 2R1, Canada}

\author{P.~Alemany}
\email{p.alemany@ub.edu}
\affiliation{Departament de Qu\'{i}mica F\'{i}sica and Institut de Qu\'{i}mica Te\`{o}rica i Computacional (IQTCUB), Universitat de Barcelona, Diagonal 647, 08028 Barcelona, Spain}

\author{E.~Canadell}
\email{canadell@icmab.es}
\affiliation{Institut de Ci\`{e}ncia de Materials de Barcelona (CSIC), Campus de la UAB, 08193 Bellaterra, Spain}

\author{K. Bechgaard}
\affiliation{Department of Chemistry, H.C. {\O}rsted Institute, Copenhagen, Denmark}

\date{\today}

\begin{abstract}
Comparing resistivity data of quasi-one dimensional superconductors \tmp6 and \tmc along the least conducting $c^\star$-axis and along the high conductivity $a$ -axis as a function of temperature and pressure, a low  temperature regime is observed  in which a unique scattering time  governs transport along both directions of these anisotropic conductors.  However, the pressure dependence of the anisotropy implies a large pressure dependence of the interlayer coupling. This is in agreement with the results of first-principles DFT calculations implying methyl group hyperconjugation in the TMTSF molecule. In this low temperature regime,  both materials exhibit for $\rho_c$ a temperature dependence $aT+bT^2$.  Taking into account the strong pressure dependence of the anisotropy, the $T$-linear $\rho_c$ is found to correlate with the suppression of the superconducting $T_c$, in close analogy with $\rho_a$ data.  This  work is revealing the domain of existence of the 3D coherent regime in the generic \tms2x phase diagram and provides further support for the correlation between $T$-linear  resistivity and superconductivity in non-conventional superconductors.
\end{abstract}


\maketitle

\section{Introduction}
\begin{figure}[h]
\centerline{\includegraphics[width=0.8\hsize]{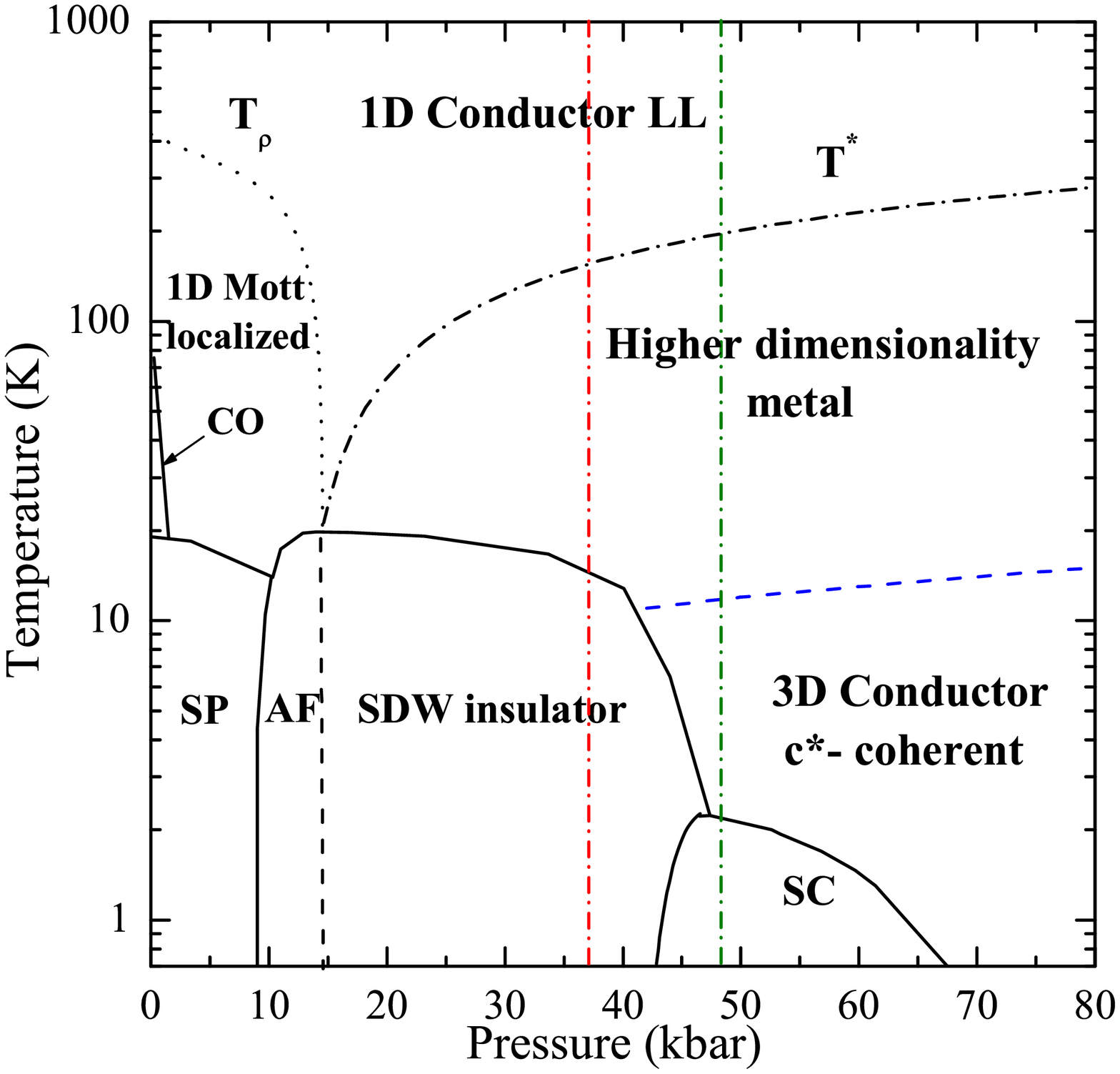}}
\caption{Generic temperature-pressure phase diagram of \tm2x. The origin of the pressure scale refers to the \tmps compound. The vertical dashed-dotted lines at 37 and 48 kbar are the estimated locations of \tmp6 and \tmc under ambient pressure, respectively. In \tmp6, the SDW order vanishes at a pressure \pc = 9.4 kbar~\cite{Kang10} whereas for \tmc this critical point is located at negative pressures. The dashed line indicates the crossover between the high-temperature quasi-1D and the low-temperature coherent regimes, as discussed in the main text.}
\label{DPPF6}
\end{figure}

As seen in Fig.~\ref{DPPF6}, the close proximity between superconductivity (SC) and an antiferromagnetic phase (AF/SDW) is a key feature of the temperature-pressure phase diagram of the \tm2x series (where TM is an electron donating organic molecule such as \tst or \tsm and X is a monoanion) of organic conductors~\cite{Jerome82,Bourbonnais08,Kang10}. This situation is observed for all members of the family with anions X = PF$_{6}$, AsF$_{6}$, ReO$_{4}$,... when the nesting of the quasi-one dimensional Fermi surfaces is destroyed under pressure near a critical pressure \pc and a non-magnetic metallic state becomes the new ground state. Because superconductivity exists mostly on the metallic side of this magnetic instability it is important to understand the nature of this metallic ground state. Recently, one of its striking feature was brought forward, namely the existence of a  temperature dependence of the longitudinal resistivity, $\rho_a$ behaving like $aT+bT^2$, at odds with the standard Fermi liquid description of metals~\cite{Doiron09,Doiron10,Doiron10a}. Furthermore, the $T$-linear contribution to the resistivity was found to be directly correlated with the superconducting $T_c$ in close analogy with cuprates~\cite{Daou09} and iron-pnictides superconductors~\cite{Fang09,Chu09}. This finding is surprising enough to warrant the performance of all possible additional confirmation  on these superconducting materials using other samples.
 
The present study reports new measurements on different samples of the transverse transport along the least conducting $c^{\star}$-axis (\textit{i.e.} normal to the $a b$ plane), $\rho_c$, and addresses the comparison between $\rho_a$ and $\rho_c$ as a function of pressure and temperature. The $c^{\star}$-axis is, in the literature, the preferred direction for most transport studies since it provides easier and more reliable measurements of the resistivity~\cite{Korin85,Moser98,Naughton88}.

Earlier works have revealed a $c^{\star}$-axis transport that goes from an insulating to a metallic temperature dependence at a temperature $T^\star$  taken as the signature of a crossover between two regimes~\cite{Moser98,Vescoli98,Giamarchi04}: a one dimensional (1D) high-temperature regime and, at low temperature, the regime of a higher dimensionality metal. The present work  focuses on the low temperature domain where a 3D anisotropic coherent band picture  prevails, in accordance with the observation of a transverse Drude edge~\cite{Henderson99} at liquid helium temperature. An important result of this investigation is the finding of an unexpectedly large pressure dependence for the interlayer coupling along $c^{\star}$, leading in turn to a significant drop of the $\rho_c/\rho_a$ anisotropy under pressure.

\section{$\rho_c/\rho_a$ anisotropy and 3D coherent regime} \label{anisotropy}

 \tmp6 and \tmc single crystals used for the $c^{\star}$-axis measurements have two contacts evaporated on both $a b$ planes and have a room temperature resistivity of 50 and 28 $\Omega$cm, respectively. These samples were measured with their $a$-axis counterpart in the same pressure cell, allowing a comparison of the temperature dependence of $\rho_a$, and $\rho_c$ at exactly the same pressure points for two different samples. Experiments were performed at eight successive pressures from 8.4 up to 20.8~kbar for \tmp6 and six successive pressures from 1.5 up to 17~kbar for \tmc . A slow cooling rate ($\leq$ 5 K/hour) was used below 50 K to ensure adequate thermalization and to optimize the anion ordering in \tmc. The experimental set up has been detailed in references~\cite{Doiron09,Doiron10}.

The main purpose of this study is to look for the influence of the nearby magnetically ordered state on the electron scattering rate in the metallic phase. The spin density wave (SDW) phase is actually the stable ground state in the phase diagram of \tmp6 up to the critical pressure \pc = 9.4~kbar~\cite{Vuletic02,Kang10} and this critical point can be approached by adequate control of the pressure. For \tmc, the conducting state is stable at ambient pressure although a magnetic phase is never far since it can be stabilized whenever the Fermi surface is left unfolded by the anion disorder~\cite{Pouget83,Takahashi82}. Consequently, the critical pressure of \tmc cannot be determined and is assumed to be negative (see the vertical lines in Fig.~\ref{DPPF6}).

In both materials, the polynomial analysis, already used for $\rho_a$ data~\cite{Doiron10,Doiron10a} and to be detailed in the next section, enables us to determine at every pressure a residual resistivity $\rho_{0c}$ in order to determine the temperature dependent inelastic scattering $\Delta\rho_{c}=\rho_{c} - \rho_{0c}$. The same quantity is also determined for $\rho_a$: $\Delta\rho_{a}=\rho_{a} - \rho_{0a}$.

\begin{figure}[h]	
\centerline{\includegraphics[width=0.6\hsize]{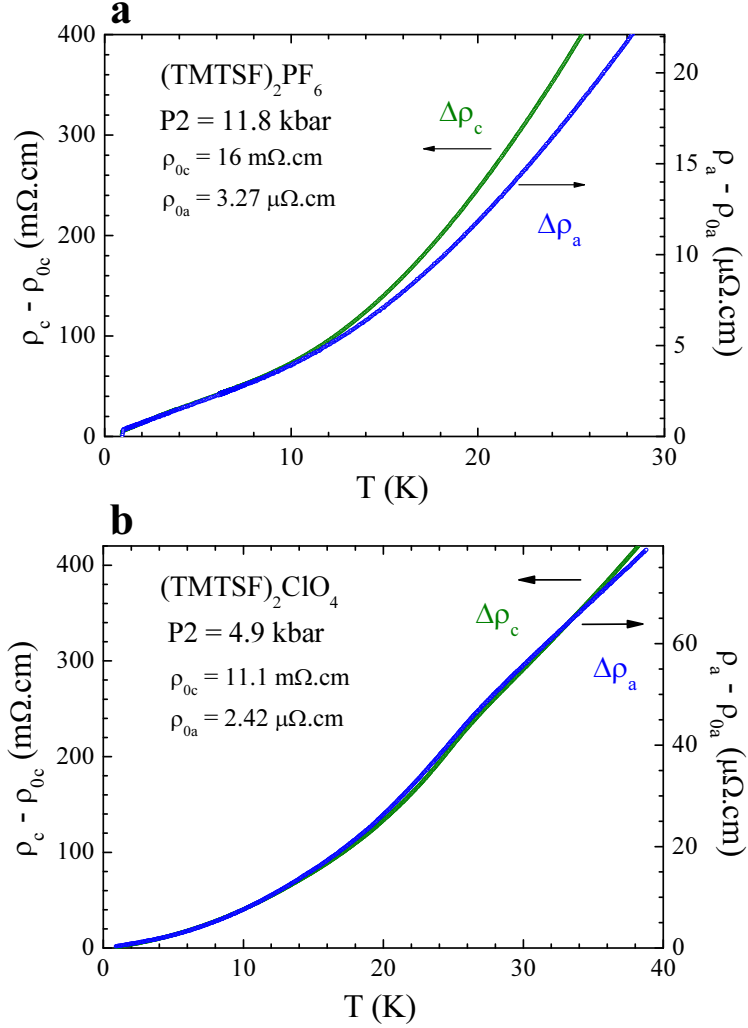}}
\caption{\textbf {a}: Temperature dependence of $\rho_{c}-\rho_{0c}$ and $\rho_{a}-\rho_{0a}$ for \tmp6 under 11.8~kbar. Similar data have been obtained at every pressure up to 20.8~kbar. These data show the onset of an increase of the anisotropy around 12 K at 11.8~kbar, temperature which increases up to 15 K for the two highest pressures. At the same time, $\Delta \rho_{c} / \Delta \rho_{a}$ at 10 K (the ratio between left and right scales) decreases from 18400 at 11.8~kbar down to 7400 at 19~kbar. \textbf {b}: Temperature dependence of $\rho_{c}-\rho_{0c}$ and $\rho_{a}-\rho_{0a}$ for \tmc under 4.9~kbar. These data show that, for \tmc at variance with \tmp6, $\Delta \rho_{c} / \Delta \rho_{a}$ at 10 K is only 5300 (the ratio between left and right scales) and that the onset of an increase of the anisotropy starts above 30 K.}
\label{fig2}
\end{figure}
Subsequently, in Fig.~\ref{fig2} we compare the inelastic contribution to the resistivity for current along the $a$ and $c^{\star}$ axes, for \tmp6 at 11.8~kbar and \tmc at 4.9~kbar. It is remarkable that both directions reveal a similar temperature dependence up to 12~K and  30~K in \tmp6 and \tmc~respectively. This behaviour allows us to define a unique scattering time at low temperature governing both components of  transport (a 3D coherent regime) and an anisotropy $\Delta \rho_{c} / \Delta \rho_{a}$ which is the ratio between left and right scales in Fig.~\ref{fig2}.

To the best of our knowledge, the upper limit for the 3D coherent regime and its pressure dependence have not yet been addressed in these quasi 1D conductors. 
 We notice on Fig.~\ref{fig2} the interesting feature that the resistance along $c^{\star}$ looks "more metallic" than the resistance along $a$ when the temperature rises above the coherent regime. This is understood in terms of the particular crossover in these 1D conductors where at high temperature $\rho_{c}$ is insulating, increasing on cooling, due to 1D physics~\cite{Moser98,Biermann01}. A metallic behaviour for $\rho_{c}$ is recovered only below the $T^{\star}$ crossover.

From our data, a fully coherent regime  prevails at a temperature below 12~K in \tmp6  when both components of the resistivity exhibit a similar temperature dependence. This upper limit for transverse coherence should be bounded by the kinetic coupling, $t_c$, along $c^{\star}$. This coupling is presumably very small compared to the coupling along the other directions. 

 The temperature domain above $t_c$ might actually correspond to the weakly-incoherent regime of 2D conductors~\cite{Kartsovnik06,Singleton07} in which Kohler's rule~\cite{Cooper86} as well as angular magnetoresistance oscillations are still observed~\cite{Kang92,Osada91,Danner94,Danner95,Sugawara06}. 

Determining the onset of the temperature dependent anisotropy at different pressures enables us to draw an estimate for  the upper limit of the temperature domain in which the $c^{\star}$-axis motion is fully coherent, as seen in Fig.~\ref{DPPF6}.

\begin{figure}[h]
\centerline{\includegraphics[width=0.8\hsize]{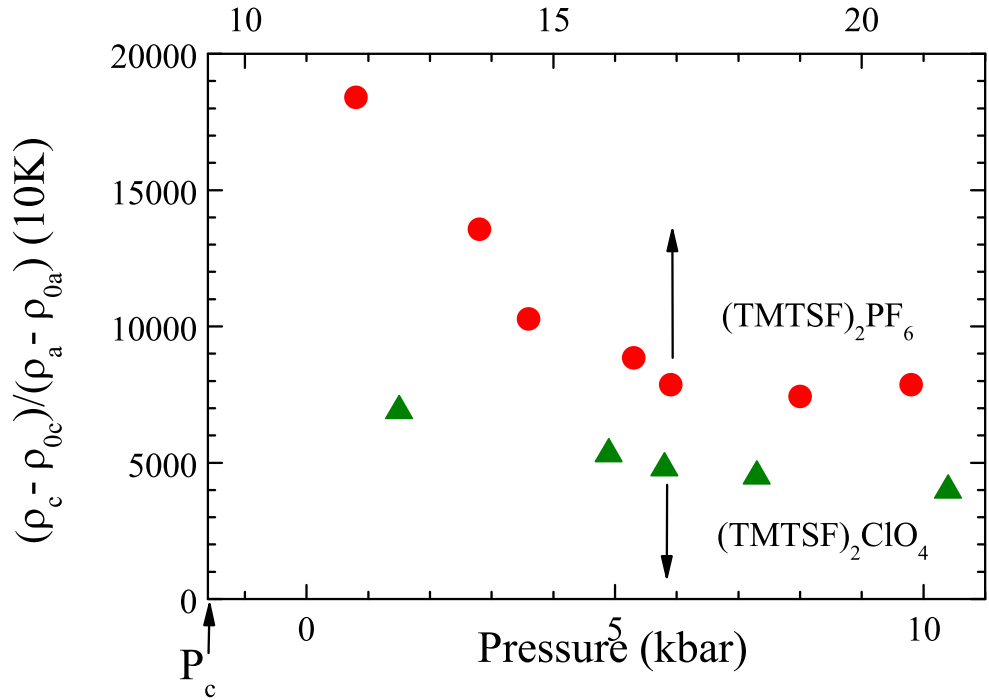}}
\caption{Pressure dependence of the anisotropy of resistivity $(\rho_{c}-\rho_{0c}) / (\rho_{a}-\rho_{0a})$ measured in the coherent regime at 10K. Data are displayed for \tmp6 (upper pressure scale) and \tmc (lower pressure scale) with a shift of 11~kbar between the two pressure scales.}
\label{fig3}
\end{figure}

In Fig.~\ref{fig3} we show the pressure dependence of the anisotropy derived from the ratio $\Delta \rho_{c} / \Delta \rho_{a}$ in the coherent regime at 10~K. We see that the  pressure dependence of the anisotropy is quite prominent in both compounds. Although we present these anisotropy data on the same figure for both compounds, it is difficult to compare absolute values obtained for \tmp6 and \tmc  since $\Delta \rho_{c} / \Delta \rho_{a}$ at 10~K depends on the absolute resistivities under ambient conditions (1~bar and 300~K). Nevertheless, the pressure dependence is reliable. 

In the case of open Fermi surfaces~\cite{Ishiguro98}, the anisotropy in the 3D coherent regime  reads, $\rho_{c} / \rho_{a} \propto(t_a/t_c)^{2}(a/c)^2$. Hence, such a large drop of the anisotropy is unexpected since a naive view could suggest the weak coupling between the $a b$ planes to be less pressure dependent than the coupling along the chain axis.

Interestingly, the coupling along $c^\star$ although quite small is affecting other physical properties which have been measured under pressure in both materials.
The unnesting parameters of the band structure, $t^{'}_{b}$ and  $t^{'}_{c}$ both play an important role in the $T-P$ and $T-P-H$ phase diagrams of \tms2x. 

First, when $t^{'}_{b}$ exceeds a critical unnesting band integral, the SDW ground state is suppressed in favour of a metallic phase with the possibility of restoration of spin density wave phases under magnetic field along $c^{\star}$ ($FISDW$ for field-induced SDW)~\cite{Ishiguro98}.
Second, the critical temperature for the stabilisation of the $FISDW$ subphases, $T_{FISDW}(H)$ should be steadily increasing from zero in a 2D conductor or in a fully nested 3D conductor in the "standard model"~\cite{Heritier84,Heritier86}.
However, since the real system is neither 2D nor perfectly nested ($t^{'}_{c}\geq $0), there exists a threshold field $H_{T}$ for the appearance of $FISDW$ subphases defined by $T_{FISDW}(H_{T})$=$t^{'}_{c}$~\cite{Montambaux86}.

Early experiments on the $FISDW$ of \tmc under pressure at 1.5~K~\cite{Creuzet85} revealed an increase of  $H_{T}$ of about 30$\%$  kbar$^{-1}$. Subsequent measurements on Bechgaard salts  under pressure performed down to very low temperature did reveal a threshold field increasing from 4.5 T at 8~kbar to 8~T at 16~kbar on \tmp6 ~\cite{Danner95} and a somewhat similar pressure dependence in \tmc~\cite{Kang93}. Such a large pressure dependence of $H_T$ implies a similarly large pressure dependence of $t^{'}_{c}$ within  the  "standard model"  with a concomitant increase of the interlayer coupling $t_{c}$.

As far as the absolute value of $t_{c}$ is concerned, not much is known besides an early calculation published in 1983 for the case of \tmtsfreo4 giving $t_{c}\approx$ 1~meV~\cite{Grant83}. In addition, an extended H\"uckel calculation has provided for \tmp6 a value of 0.8~meV for $t_{c}$~\cite{Balicas94}.
Given the observed large pressure dependence of the $c^\star$ coupling it is therefore important to see whether this pressure dependence can be explained by the pressure-induced deformation of the  band structure.

\section{DFT calculation}

First-principles calculations were carried out for \tmp6 for which reliable structural data have been obtained under 1 bar~\cite{ThoRin1981} and 6.5 kbar~\cite{GalGau1986}. We used a numerical atomic orbitals DFT approach~\cite{HohKoh1964,KohSha1965} developed for efficient calculations in large systems and implemented in the SIESTA code~\cite{SolArt2002}. The generalized gradient approximation to DFT and, in particular, the functional of Perdew, Burke and Ernzerhof was adopted~\cite{PBE96}. Only the valence electrons are considered in the calculation, with the core being replaced by norm-conserving scalar relativistic pseudopotentials~\cite{tro91} factorized in the Kleinman-Bylander form~\cite{klby82}. We have used a split-valence double-$\zeta$  basis set including polarization orbitals as obtained with an energy shift of 10 meV for all atoms~\cite{arsan99}. The energy cutoff of the real space integration mesh was 250 Ry and the Brillouin zone was sampled using grids of (4$\times $4$\times $4) k-points~\cite{MonPac76}. 

\begin{figure}[h,t]
\centerline{\includegraphics[width=0.8\hsize]{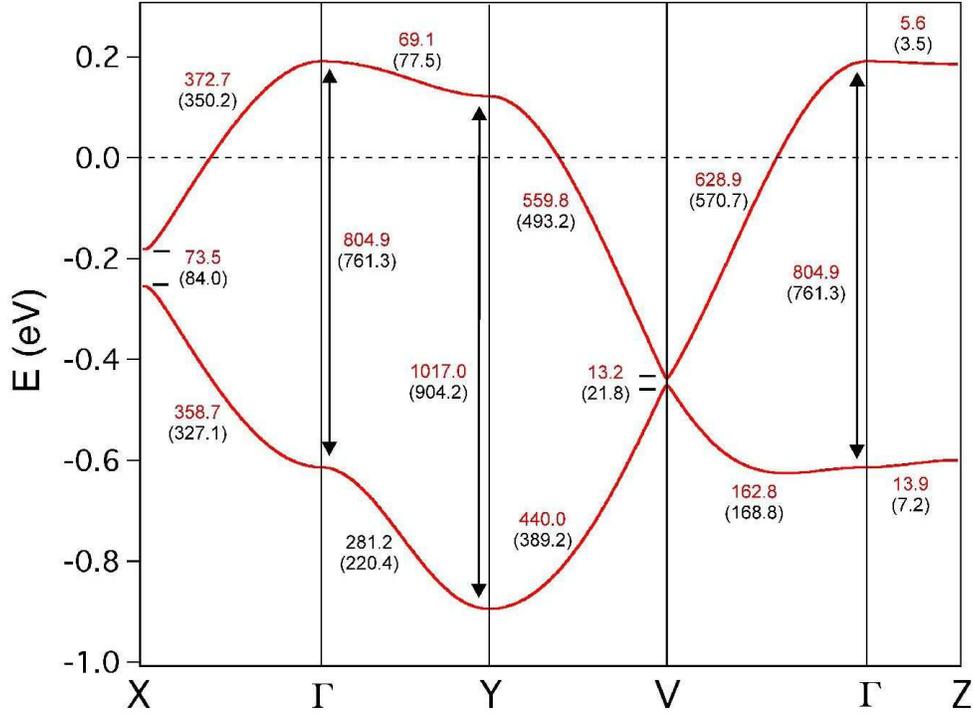}}
\caption{Calculated band structure for \tmp6 using the crystal structure obtained under 6.5 kbar. The bandwidths and gaps reported are for the structure under 6.5 kbar whereas the data in parenthesis correspond to the 1 bar structure. All values are given in meV. The dashed line refers to the Fermi level and $\Gamma $ = (0, 0, 0), X = (1/2, 0, 0), Y = (0, 1/2, 0), V = (1/2, 1/2, 0) and Z = (0, 0, 1/2) in units of the monoclinic reciprocal lattice vectors.}
\label{Band_dispersion}
\end{figure}

The calculated band structure under 6.5 kbar is reported on Fig.~\ref{Band_dispersion}. The main parameters of the band structure at both 6.5 kbar and 1 bar are also given in that figure. Using the full band dispersion at the $\Gamma$ point we obtain for $t_{a}$ an increase of 0.88 \% per kbar. Although to the best of our knowledge there is no direct experimental data for comparison, this value matches well a previous more qualitative estimation by Ducasse et al.~\cite{DucAbd1986}, 0.75 \%. Concerning the effective transverse interaction $t_{b}$ let us note that taking the values of Fig.~\ref{Band_dispersion} for the $\Gamma \rightarrow $ Y line it looks as if $t_{b}$ was decreasing from 1 bar to 6.5 kbar, something not easily matching the idea that the nesting of the Fermi surface deteriorates under pressure leading to the suppression of the SDW instability. However when the full Brillouin zone is explored it is found that when moving from the $\Gamma \rightarrow $ Y line there is progressive change which quite soon results with an inversion of this behaviour. In particular, all along the Fermi surface the effective transverse interaction increases under pressure. Thus, the DFT band structure of \tmp6 seems to capture well the essential features of its pressure dependence. Let us note that the same type of calculations has already provided interlayer dispersion values consistent with experimental results for other molecular metals like $\alpha $-(BEDT-TTF)$_2$KHg(SCN)$_4$~\cite{FouPou2010} and $\beta $-(BEDT-TTF)$_2$I$_3$~\cite{LeeNie2003}.  

Turning to a comparison between anisotropy data displayed on Fig.~\ref{fig3} and theory we note that the results on Fig.~\ref{fig3} provide a drop of the anisotropy by a factor$ \approx$ 2.6 between 1~bar and 6.5~kbar. Using the full band dispersion at the $\Gamma$ point and the dispersion along the $\Gamma$-Z direction we notice that the square of the ratio of dispersions along $a$ and $c$ is dropping by a factor 2.3 under 6.5 kbar. This is admittedly close to the experimental drop of 2.6 on Fig.~\ref{fig3}. Consequently, the DFT calculation of the band structure under pressure supports the unexpected strong dependence of the anisotropy. 

The origin of this result lies in the well known ability of methyl groups to propagate the $\pi $-type delocalisation (hyperconjugation) through its $\pi _{CH_3}$ and $\pi _{CH_3}^*$ orbitals~\cite{AlBuWh1985}. Thus, even if weakly, the HOMO of \tsm extends towards the outer methyl groups. In the crystal structure of \tmp6 there are three short direct \tsm interactions per dimer along the $c$ direction which implicate these methyl groups. These contacts become shorter under pressure. For instance the C$\cdot \cdot \cdot $C distances are 3.890, 3.890 and 3.971 \AA~at 1 bar and become 3.705, 3.705 and 3.936 \AA~at 6.5 kbar. In that way, the interlayer HOMO$\cdot \cdot $HOMO interactions increase. Even if in absolute terms the effect is small, the inherent weakness of the interaction along $c$ magnifies the variation and leads to the drop in the calculated values and in Fig.~\ref{fig3}.

Assuming the unnesting coupling along $c$ to be given by  $t^{'}_{c}= t^{2}_c/t_a$, the order of magnitude for its pressure dependence derived from the calculation amounts to 20$\% kbar^{-1}$. This is admittedly  in fair agreement with the observed strong pressure dependence of the FISDW onset field  for both \tmc and \tmp6, \textit{vide supra}.

\section{Correlation between $\rho_c$ and \tc}

We shall now develop the procedure used to analyse the temperature dependence of the transverse resistivity,  $\rho_c$, from the raw experimental data. 

\begin{figure}[h]
\centerline{\includegraphics[width=0.7\hsize]{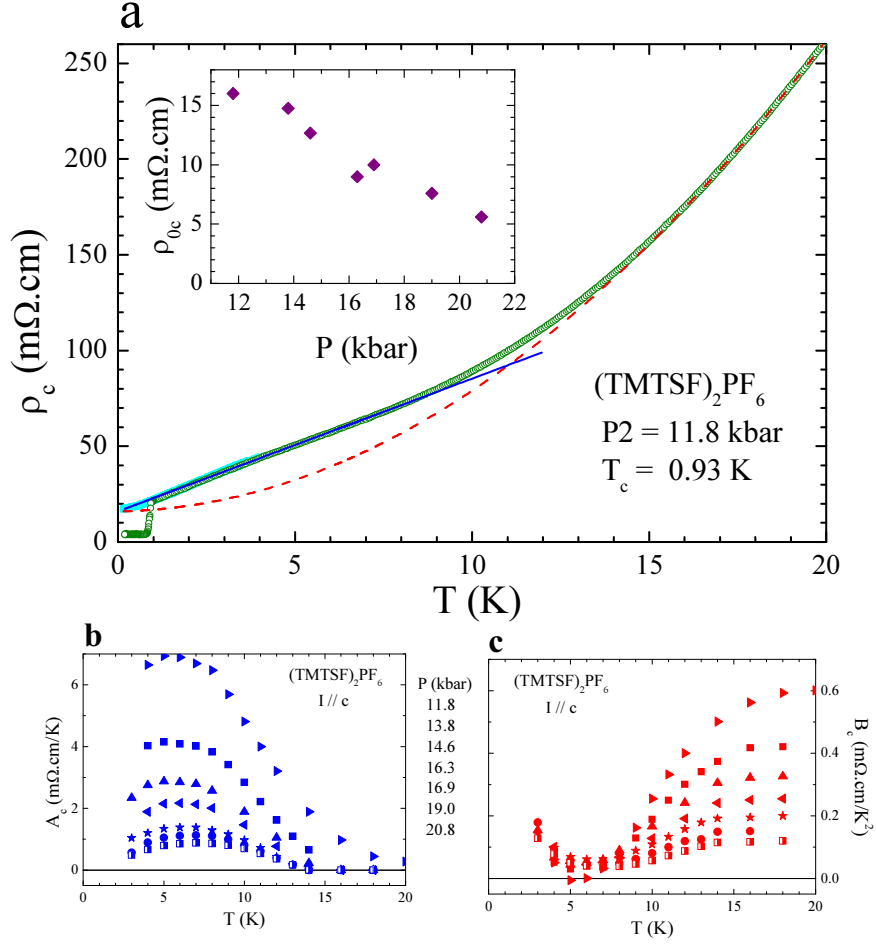}}
\caption{\textbf{a}: $c^\star$-axis resistivity $\rho_c$ of \tmp6 at 11.8 kbar versus temperature, at zero field and under $H=0.05$ T applied along $c^*$ in order to suppress superconductivity. The second order polynomial fit, $\rho_c(T) = \rho_{0,c} +  A_c(T) T + B_c(T) T^2$, according to the sliding fit procedure described in the text is shown for the $T$ intervals $(3-7)$ K (blue) and $(18-22)$ K (dashed red). The insert displays the pressure dependence of the residual resistivity derived from the lowest temperature fit (see text). Temperature dependence of $A_c$ (\textbf{b}) and $B_c$ (\textbf{c}) at different pressures as indicated. Every temperature point corresponds to the center of the $4$K window used for the fit.}
\label{RTPF6}
\end{figure}

The $\rho_c$ data on \tmp6 at a pressure of 11.8 kbar, closest to \pc in our experiment, are displayed in Fig.~\ref{RTPF6}(a) up to 20~K. We see that the resistivity can be analysed by the sum of an elastic contribution plus inelastic linear and quadratic contributions such that $\rho_c(T) = \rho_0 +A_{c}T + B_{c}T^2$.  It is clear from the data shown in Fig.~\ref{RTPF6}(a) that the relative weight of $A_c$ and $B_c$ is indeed changing with temperature, with $A_c$ and $B_c$ being dominant at low and high temperatures respectively. A pure linear resistivity is observed at that pressure  for the  $c^{\star}$-axis transport below about 8~K, and down to about 0.3~K by using a weak magnetic field of $H=0.05$~T along $c^{\star}$ to suppress superconductivity. Above this linear regime, at about 15~K and above, the resistivity is quadratic in temperature, as indicated by the dashed red line in Fig.~\ref{RTPF6}(a). 

\begin{figure}[t]
\centerline{\includegraphics[width=0.7\hsize]{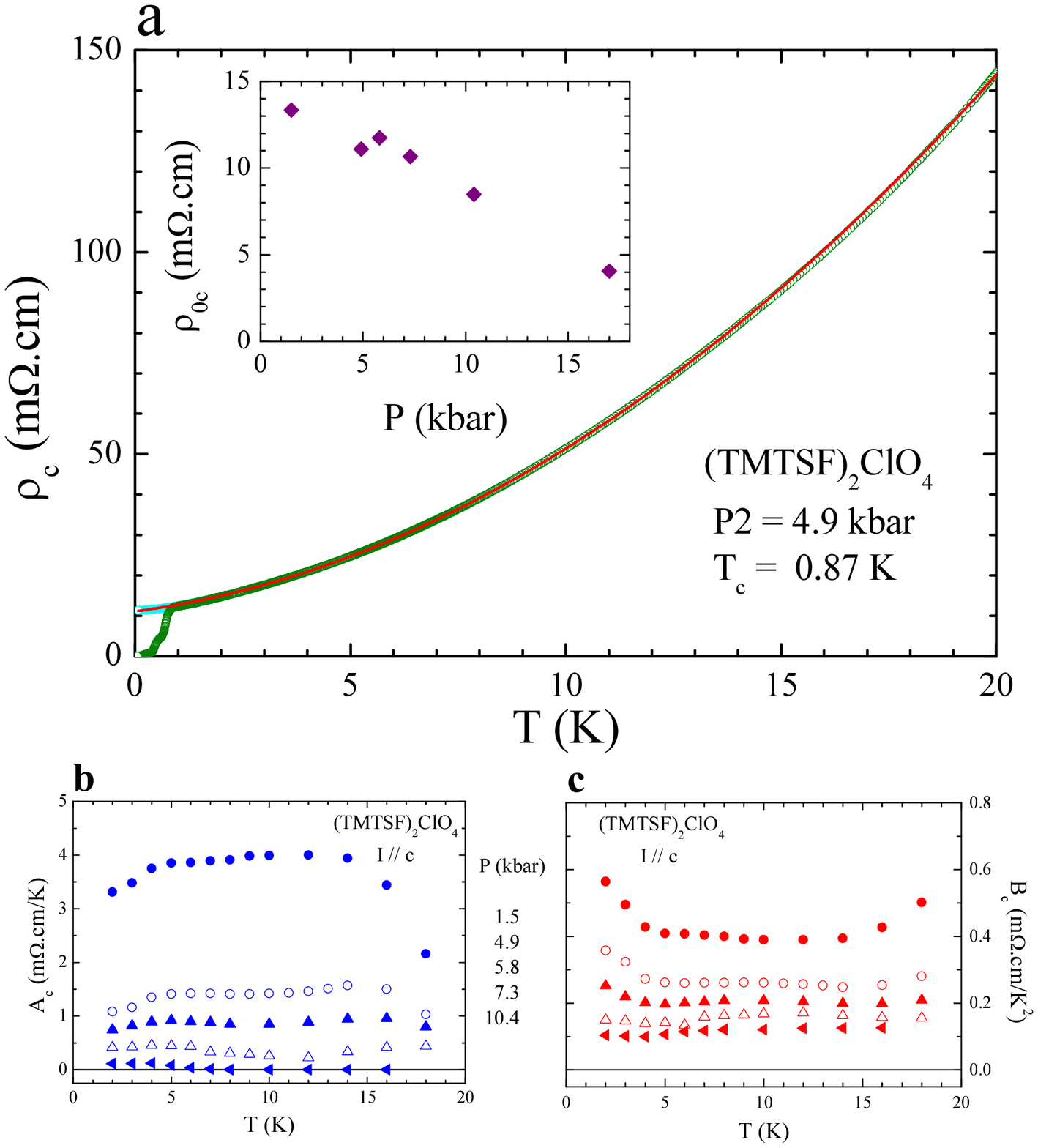}}
\caption{\textbf {a}: $c^\star$-axis resistivity $\rho_c$ of \tmc at 4.9~kbar versus temperature, at zero field and under $H=0.05$ T applied along $c^{\star}$ in order to suppress superconductivity. The second order polynomial fit, $\rho_c(T) = \rho_{0,c} +  A_{c}(T) T + B_{c}(T) T^2$ described in the text is shown in red for the $T$ interval (8 - 12)~K. The insert displays the pressure dependence of the residual resistivity derived from the lowest temperature fit (see text). Temperature dependence of $A_c$ (\textbf{b}) and $B_c$ (\textbf{c}) at different pressures as indicated. Every temperature point corresponds to the center of the 4~K window used for the fit.}
\label{RTClO4}
\end{figure}

As far as \tmc is concerned, see Fig.\ref{RTClO4}, the same polynomial analysis can be performed but it is more difficult to distinguish the purely linear or purely quadratic regimes. As shown on resistivity data at 4.9~kbar displayed in Fig.~\ref{RTClO4}(a), both inelastic contributions are coexisting over the entire temperature domain. 
In order to capture the evolution of $A_c$ and $B_c$ over the entire pressure and temperature range examined, we use the same sliding fit procedure employed in the context of in-chain data~\cite{Doiron10,Doiron10a}, whereby we fit the resistivity curves to $\rho_{c}(T) = \rho_{0,c} +A_{c}T + B_{c}T^2$ over a sliding temperature window of 4~K. This fitting procedure has been carried out at all pressures keeping the value for the residual resistivity constant for all fits performed at a given pressure ($\rho_{0,c}$ is determined by the fit for the lowest temperature window).The result of this analysis on \tmp6 for all our measured pressures is shown in the bottom panels of Fig.~\ref{RTPF6} where the existence of a low-temperature linear regime and a more quadratic high-temperature regime is clear. Turning to the \tmc data, this decomposition of the resistivity gives an excellent fit to the data over a large temperature range up to the anion ordering temperature with only a small variation of the fit parameters, as shown in the bottom panels of Fig.~\ref{RTClO4}.

The sliding fit procedure gives nearly temperature independent prefactors for \tmc, but a strong temperature dependence is noticed in \tmp6, especially at the lowest pressures. The difference between the data for both compounds may be ascribed to different distances from  the critical point \pc. As a result, a stronger linear term can be anticipated in \tmp6 which is closer to \pc than \tmc, if the amplitude of the linear contribution is related to the proximity of the magnetic ground state. Moreover, \tmc exhibits a folded Fermi surface which is likely to interfere with the development of the linear contribution as obtained in \tmp6. At any rate, the present study of transport in the metallic phase of \tmp6 and \tmc along the least conducting direction shows that the scattering rate comprises linear and quadratic terms, as seen for transport along the chains.

However, given the pressure dependence of the anisotropy displayed on Fig.~\ref{fig3} which is derived from the anisotropy of the inelastic scattering at low temperature one could expect the residual resistivity to show a similar effect. According to the inset of Figs.~\ref{RTPF6}(a) and \ref{RTClO4}(a) the pressure dependence of the residual resistivity is significantly larger than that of the inelastic contribution. This feature can be understood as being because the residual resistance is quite sensitive to defects and was always found in the measurements of several samples less reliable than the temperature dependent resistance. 
  
The decomposition of the inelastic scattering  term as the sum of linear and quadratic terms rather than a power law suggests that a regular Fermi liquid scattering channel is superimposed on a more unusual one, the latter being most likely connected to the scattering on low energy spin fluctuations. It is worth noting that in the context of high-\tc cuprates, such superimposed scattering channels seems to give the best description of the normal-state resistivity data, such as reported on Tl$_2$Ba$_2$CuO$_{6+\delta}$~\cite{Mackenzie96,Proust02} and La$_{2-x}$Sr$_x$CuO$_4$~\cite{Cooper09}. It does not, however, necessarily require a `two-fluid' like separation of the carriers (hot and cold regions on the Fermi surface for instance) as it can take place for one type of carriers when these are coupled to a wide fluctuation spectrum.
 
This has been indeed shown by scaling theory for the calculation of the electron-electron scattering rate close to  SDW ordering in a quasi-1D metal (the results are summarized in  reference~\cite{Doiron10}). Near the critical pressure, where SDW connects with superconductivity, spin fluctuations are strong and their spectrum is sharply peaked at very low energy ($\omega_{sf}$), which is comparable to or smaller than temperature $T$ (see, \textit{e.g.}, reference~\cite{Bourbonnais09}). Under these conditions, their contribution yields a clear linear  temperature dependence for the scattering rate, a known result for electrons interacting with low-energy  bosonic spin modes  in {\it two} dimensions (see e.g.,~\cite{Abanov03}). Moving away from critical pressure, spin fluctuations  decrease, their spectral peak widens, drops in amplitude and gradually moves  to much higher energy (an evolution confirmed on experimental grounds by NMR spin-lattice relaxation rate under pressure in the Bechgaard salts~\cite{Creuzet87b,Brown08}). This corresponds to an intermediate situation where electrons scatter on both low and sizable energy modes.  The former  modes are still responsible for a linear term, though with a decreasing amplitude under pressure, while the latter modes favor  the opening of a  different scattering channel  at high energy which fulfills the Fermi liquid requirements ($\omega_{sf}\gg T$).   Scaling theory calculations  confirm that as one moves away from the critical pressure, the scattering rate is no longer perfectly linear in temperature above \tc,  but develops some curvature that is fitted quite satisfactorily by a $aT + bT^2$ form (see Fig. 10 of reference~\cite{Doiron10}).
\begin{figure}[t]
\centerline{\includegraphics[width=0.8\hsize]{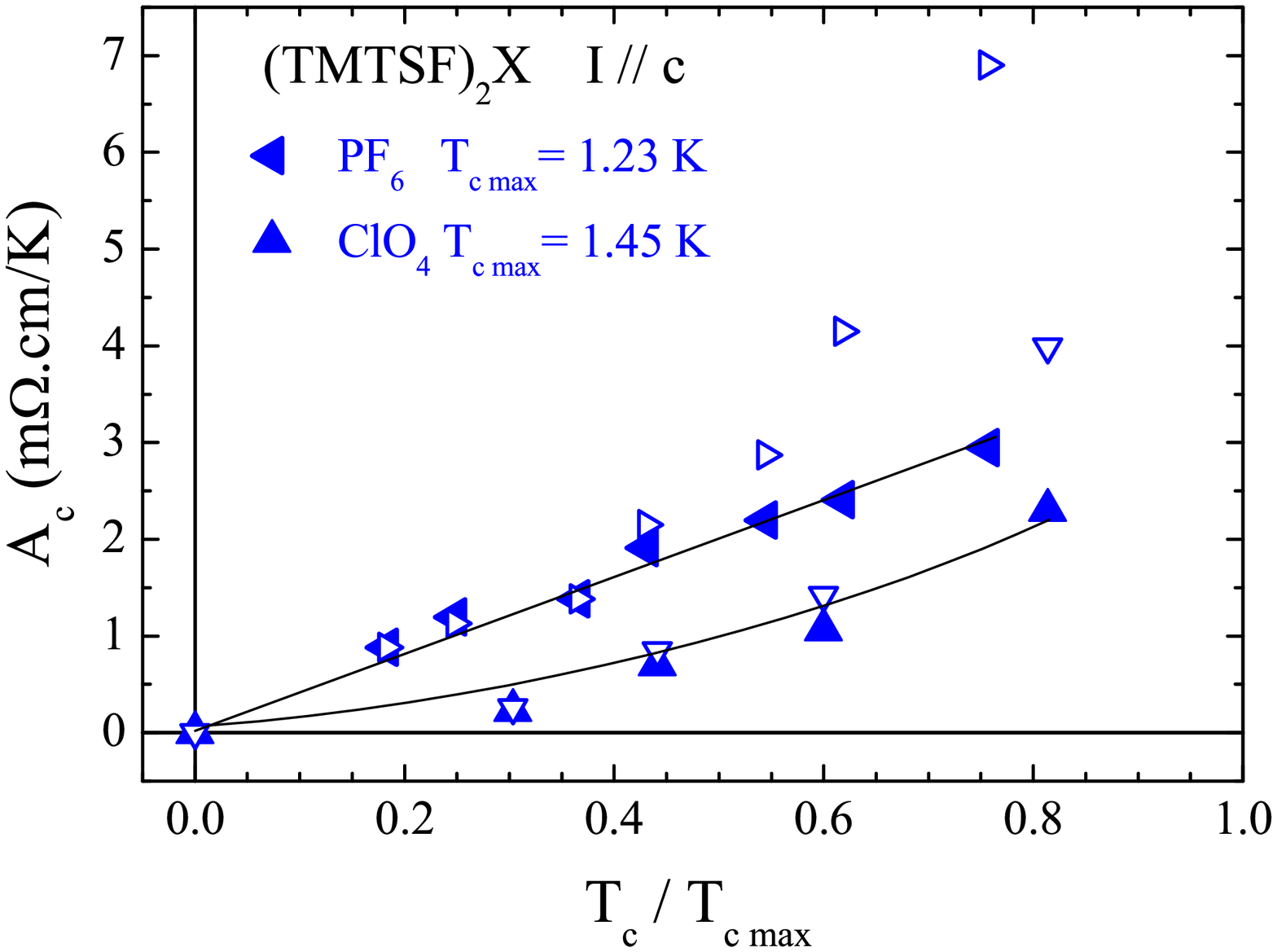}}
\caption{$A_c$ coefficient versus reduced \tc in \tmp6  and \tmc ; empty symbols are the raw data for $A_c$ determined at the temperature corresponding to its maximum value, namely, T = 5~K for \tmp6 and 10~K for \tmc ; full symbols are the $A_c$ values corrected for the pressure dependence of the anisotropy (see text). The maximum \tc  for \tmp6 is the value obtained at 8.4~kbar on the same sample, a pressure which is located in the inhomogenous SDW/metal state. The maximum \tc for \tmc comes from $\rho_{c}$ data at 1~bar obtained by S. Yonezawa~\cite{Yonezawa08} on a very slowly cooled sample from the same batch.}  
\label{ATc}
\end{figure}

We have plotted in Fig.~\ref{ATc}, the coefficient of the T-linear contribution, $A_c$, versus the reduced \tc ($T_c / T_{c max}$) for both compounds.
Given that a significant contribution to the drop of $A_c$  under pressure is actually due to the decrease of the anisotropy, it is of interest to plot the $A_c$ coefficient corrected for the pressure-dependent anisotropy. In order to correct for this extrinsic drop of anisotropy under pressure we have divided the raw $A_c$ values by the ratio of the anisotropy at each pressure point to the anisotropy at the highest pressure for each material (20.8~kbar for \tmp6 and 10.4~kbar for \tmc). We have neglected the pressure dependence of the band parameters given the very small variation of $\rho_{0a}$ measured at the same time and shown in Fig.4 of ref~\cite{Doiron10}. 

The result of this procedure, also plotted  in Fig.~\ref{ATc}, makes the dependence of $A_c$ on \tc   quasi-linear. This behaviour is in qualitative agreement with the RG theory~\cite{Doiron10}. The present experiments do not approach the region very close to \pc (or the highest \tc) where a further  enhancement of $A_c$ \textit{albeit} non diverging is expected according to the one loop RG  theory~\cite{Doiron10}. 

The  vanishing  of superconductivity of \tmc above 8 kbar is likely due to the remnence of defects related to an incomplete anion ordering. Such a  vanishing is not observed in \tmp6 which is expected to be a cleaner superconductor. Hence, superconductivity in \tmp6  persists up to the highest pressure of our study.

\section{Conclusion}
In summary, the investigation of the metallic region of the \tms2x phase diagram using the pressure and temperature dependence of the transverse resistivity $\rho_c$ reveals several new  features.

First, a comparison between $\rho_a$ and $\rho_c$ defines a domain of existence for a band-like motion of carriers along $c^{\star}$, namely below 12~K or so for \tmp6 and up to 30~K for \tmc, with a single scattering rate governing the temperature dependence of transport along $a$ and $c^{\star}$ allowing a mapping of the 3D coherent regime.

Second, the anisotropy of resistivity in the 3D coherent regime reveals a strong  pressure dependence which suggests a pressure dependence of the coupling much stronger  along $c^\star$ than along $a$. Such a feature is actually  in agreement with the pressure dependence of the FISDW phase diagram. This experimental behaviour is fairly well accounted for by  the  DFT band structure calculation performed according to the 1~bar and  6.5~kbar structures. The origin of the strong pressure dependence of the coupling along $c^\star$ lies in the well known ability of methyl groups to propagate the $\pi $-type delocalisation (hyperconjugation) through its $\pi _{CH_3}$ and $\pi _{CH_3}^*$ orbitals.

Third, $\rho_c$ has a temperature dependence departing from the canonical Fermi behaviour since a fit such as $\rho_0 +A_{c}T + B_{c}T^2$  provides a good description of the low temperature data, in contrast to the Fermi liquid $T^2$ law. When the  pressure dependence of the anisotropy is  taken into account the relation between  $A_c$  and \tc is similar to the relation found between $A_a$ and \tc in fair agreement with the RG one loop theory~\cite{Doiron10}. 

This  work reinforces further the intimate connection between the two phenomena, also observed in cuprate and iron-pnictide high temperature superconductors~\cite{Doiron09,Taillefer10}, suggesting that it is an essential ingredient for our understanding of these materials.

This work has been supported by NSERC (Canada), FQRNT (Qu\'ebec), CFI (Canada), a Canada Research Chair (L.T.), the Canadian Institute for Advanced Research, CNRS (France) and DGI-Spain (Grants No. CSD2007-00041, FIS2009-12721-C04-03 and CTQ2008-06670-C02-02/BQU). We thank S.Yonezawa for communicating ambient pressure data of a run on \tmc performed at Kyoto.


\begin{thebibliography}{0}
\expandafter\ifx\csname natexlab\endcsname\relax\def\natexlab#1{#1}\fi
\expandafter\ifx\csname bibnamefont\endcsname\relax
  \def\bibnamefont#1{#1}\fi
\expandafter\ifx\csname bibfnamefont\endcsname\relax
  \def\bibfnamefont#1{#1}\fi
\expandafter\ifx\csname citenamefont\endcsname\relax
  \def\citenamefont#1{#1}\fi
\expandafter\ifx\csname url\endcsname\relax
  \def\url#1{\texttt{#1}}\fi
\expandafter\ifx\csname urlprefix\endcsname\relax\def\urlprefix{URL }\fi
\providecommand{\bibinfo}[2]{#2}
\providecommand{\eprint}[2][]{\url{#2}}

\end{thebibliography}


\section*{References}
\begin{thebibliography}{10}

\bibitem{Jerome82}
D.~J\'erome and H.~J. Schulz.
\newblock {\em Adv in Physics}, 31:299, 1982.

\bibitem{Bourbonnais08}
C.~Bourbonnais and D.~J\'erome.
\newblock {\em The Physics of Organic Superconductors and Conductors}, page 357.
\newblock A.~Lebed editors, Springer Verlag, Heidelberg, 2008.

\bibitem{Kang10}
N.~Kang, B.~Salameh, P.~Auban-Senzier, D.~J\'erome, C.~R. Pasquier, and S.~Brazovskii.
\newblock {\em Phys. Rev.B}, 81:100509, 2010.
\newblock cond-mat:1002.3767v1.

\bibitem{Doiron09}
N.~Doiron-Leyraud, P.~Auban-Senzier, S.~Ren\'e de~Cotret, C.~Bourbonnais,
  D.~J\'erome, K.~Bechgaard, and L.~Taillefer.
\newblock {\em Phys. Rev. B}, 80:214531, 2009.

\bibitem{Doiron10}
N.~Doiron-Leyraud, P.~Auban-Senzier, S.~Ren\'e de~Cotret, C.~Bourbonnais,
  D.~J\'erome, K.~Bechgaard, and L.~Taillefer.
\newblock {\em Eur. Phys. Jour.B}, 78:23, 2010.
\newblock DOI 10.1140/epjb/e2010-10571-4.

\bibitem{Doiron10a}
N.~Doiron-Leyraud, P.~Auban-Senzier, S.~Ren\'e de~Cotret, C.~Bourbonnais,
  D.~J\'erome, K.~Bechgaard, and L.~Taillefer.
\newblock {\em Physica B} 405:S265, 2010.
\newblock Proceedings of ISCOM 2009,
\newblock arXiv.org:0912.20492010.

\bibitem{Daou09}
R.~Daou, N.~Doiron-Leyraud, D.~LeBoeuf, S.~Y. Li,
F.~Lalibert\'e, O.~Cyr-Choini\u ere, Y.~J. Jo, L.~Balicas,
J.-Q. Yan, J.-S. Zhou, J.~B. Goodenough, and L.~Taillefer.
\newblock {\em Nat. Phys}, 5:31, 2009.

\bibitem{Fang09}
L.~Fang, H.~Luo, P.~Cheng, Z.~Wang, Y.~Jia, G.~Mu, B.~Shen, I.~I.
Mazin, L.~Shan, C.~Ren and H.~H. Wen.
\newblock {\em Phys. Rev. B}, 80:140508(R), 2009.
\newblock arXiv.org:0903.2418.

\bibitem{Chu09}
J.~H. Chu, J.~G. Analytis, C.~Kucharczyk,  and I.~R. Fisher.
\newblock {\em Phys. Rev. B}, 79:014506--1, 2009.

\bibitem{Korin85}
B.~Korin-Hamzi\'c, F.~Forro, and J.~R. Cooper.
\newblock {\em Mol. Cryst. Liq. Cryst}, 119:135, 1985.

\bibitem{Moser98}
J.~Moser, M.~Gabay, P.~Auban-Senzier, D.~J\'erome, K.~Bechgaard, and J.~M. Fabre.
\newblock {\em Eur. Phys. Jour. B}, 1:39, 1998.

\bibitem{Naughton88}
M.~J. Naughton, R.~V. Chamberlin, P.~M.~Chaikin X.~Yan, S.Y. Hsu, L.Y. Chiang,  and M.Y. Azbel.
\newblock {\em Phys. Rev. Lett.}, 61:621, 1995.

\bibitem{Vescoli98}
V.~Vescoli, L.~Degiorgi, W.~Henderson, G.~Gr\"uner, K.~P. Starkey, and L.~K. Montgomery.
\newblock {\em Science}, 281:1181, 1998.

\bibitem{Giamarchi04}
T.~Giamarchi.
\newblock {\em Quantum Physics in One-Dimension}.
\newblock Clarendon Press, Oxford, 2004.

\bibitem{Henderson99}
W.~Henderson, V.~Vescoli, P.~Tran, L.~Degiorgi, and G.~Gr\"uner.
\newblock {\em Eur. Phys. Jour. B}, 11:365, 1999.

\bibitem{Vuletic02}
T.~Vuleti\'c, P.~Auban-Senzier, C.~Pasquier, S.~Tomi\'c, D.~J\'erome,
  M.~H\'eritier, and K.~Bechgaard.
\newblock {\em Eur. Phys. J. B}, 25:319, 2002.

\bibitem{Pouget83}
J.~P. Pouget, G.~Shirane, K.~Bechgaard, and J.~M. Fabre.
\newblock {\em Phys. Rev. B}, 27:5203, 1983.

\bibitem{Takahashi82}
T.~Takahashi, K.~Bechgaard, and D.~J\'erome.
\newblock {\em J. Physique. Lett}, 43:L565, 1982.

\bibitem{Biermann01}
S.~Biermann, A.~Georges, A.~Lichtenstein, and T.~Giamarchi.
\newblock {\em Phys. Rev. Lett.}, 87:276405, 2001.

\bibitem{Kartsovnik06}
M.~V. Kartsovnik, D.~Andres, S.~V. Simonov, W.~Biberacher, I.~Sheikin, N.~D. Kushch and  H.~M\"uller.
\newblock {\em Phys. Rev. Lett}, 96:166601, 2006.

\bibitem{Singleton07}
J.~Singleton, P.~A. Goddard, A.~Ardavan, A.~I. Coldea, S.~J. Blundell, R.~D. McDonald,  S.~Tozer and J.~A. Schlueter.
\newblock {\em Phys. Rev. Lett}, 99:027004, 2007.

\bibitem{Cooper86}
J.~R. Cooper, L.~Forr\'o, B.~Korin-Hamzi\'c, K.~Bechgaard, and A.~Moradpour.
\newblock {\em Phys. Rev. B}, 33:6810, 1986.

\bibitem{Kang92}
W.~Kang, S.~T. Hannahs, and P.~M. Chaikin.
\newblock {\em Phys. Rev. Lett}, 69:2827, 1992.

\bibitem{Osada91}
T.~Osada, A.~Kawasumi, S.~Kagoshima, N.~Miura and G.~Saito.
\newblock {\em Phys. Rev. Lett}, 66:1525, 1991.

\bibitem{Danner94}
G.~M. Danner, W.~Kang, and P.~M. Chaikin.
\newblock {\em Phys. Rev. Lett}, 72:3714, 1994.

\bibitem{Danner95}
G.~M. Danner, and P.~M. Chaikin.
\newblock {\em Phys. Rev. Lett}, 75:4690, 1995.

\bibitem{Sugawara06}
S.~Sugawara, T.~Ueno, Y.~Kawasugi, N.~Tajima, Y.~Nishio and K.~Kajita.
\newblock {\em J. Phys. Soc. Jpn.}, 75:053704, 2006.

\bibitem{Ishiguro98}
T.~Ishiguro, K.~Yamaji, and G.~Saito.
\newblock {\em Organic Superconductors}.
\newblock Springer, Berlin, 1998.

\bibitem{Heritier84}
M.~H\'eritier, G.~Montambaux, and P.~Lederer.
\newblock {\em J. Phys. (Paris) Lett.}, 45:L943, 1984.

\bibitem{Heritier86}
M.~H\'eritier.
\newblock {\em Low-Dimensional Conductors and Superconductors}, page 243.
\newblock D.~J\'erome and L.~G. Caron, editors, Plenum Press, New York, 1987. 

\bibitem{Montambaux86}
G.~Montambaux.
\newblock {\em Low-Dimensional Conductors and Superconductors}, page 233.
\newblock D.~J\'erome and L.~G. Caron, editors, Plenum Press, New York, 1987. 

\bibitem{Creuzet85}
F.~Creuzet, D.~J\'erome, and A.~Moradpour.
\newblock {\em Mol. Cryst. Liq. Cryst.}, 119:297, 1985.

\bibitem{Kang93}
W.~Kang, S.~T. Hannahs, and P.~M. Chaikin.
\newblock {\em Phys. Rev. Lett}, 70:3091, 1993.

\bibitem{Grant83}
P.~M. Grant.
\newblock {\em J. Phys. (Paris) Coll.}, 44:847, 1983.

\bibitem{Balicas94}
L.~Balicas, K.~Behnia, W.~Kang, E.~Canadell, P.~Auban-Senzier, D.~J\'erome, M.~Ribault, and J.~M. Fabre.
\newblock {\em J. Phys. I (France)}, 4:1539, 1994.

\bibitem{ThoRin1981}
N.~Thorup, G.~Rindorf, H.~Soling and K.~Bechgaard.
\newblock {\em Acta. Cryst.} B37:1236, 1981.
  
\bibitem{GalGau1986}
B.~Gallois, J.~Gaultier, C.~Hauw, T.-d.~Lamcharfi and A.~Filhol.
\newblock {\em Acta. Cryst.} B42:564, 1986.

\newblock { A crystal structure under 9.8~kbar, which is however of somewhat lesser quality, was also reported in this reference and used in our study. However, when we compared the band structures at 1~bar, 6.5~kbar and 9.8~kbar we detected a somewhat erratic behaviour for the HOMO bands of the last one in several parts of the Brillouin zone. Thus we concluded that the 9.8~kbar structure although correctly describing most of the structural aspects is not precise enough for the fine description of the band structure parameters. }

\bibitem{HohKoh1964}
P. Hohenberg and W. Kohn.
\newblock {\em Phys. Rev. B} 136:864, 1964.

\bibitem{KohSha1965}
W. Kohn and L.~J. Sham.
\newblock {\em Phys. Rev. A} 140:1133, 1965.
  
\bibitem{SolArt2002}
J.~M. Soler, E. Artacho, J.~Gale, A.~Garc\'{i}a, J.~Junquera, P.~Ordej\'{o}n and D.~S\'{a}nchez-Portal.
\newblock {\em J. Phys.: Condens. Matter} 14:2745, 2002.   
 
\bibitem{PBE96}
J.~P. Perdew, K. Burke, and M. Ernzerhof.
\newblock {\em Phys. Rev. Lett.} 77:3865, 1996.  

\bibitem{tro91}
N. Troullier and J.~L. Martins.
\newblock {\em Phys. Rev. B} 43:1993, 1991.
  
\bibitem{klby82}
L. Kleinman and D.~M. Bylander.
\newblock {\em Phys. Rev. Lett.} 48:1425, 1982.

\bibitem{arsan99}
E. Artacho, D. S\'{a}nchez-Portal, P.~Ordej\'{o}n, A.~Garc\'{i}a, and J.~M. Soler.
\newblock {\em Phys. Stat. Sol. (b)} 215:809, 1999.

\bibitem{MonPac76}
H.~J. Monkhorst, and J.~D. Pack.
\newblock {\em Phys. Rev. B} 13:5188, 1976. 

\bibitem{DucAbd1986}
L.~Ducasse, M.~Abderrabba, J.~Hoarau, B.~Gallois, and J.~Gaultier.
\newblock {\em J. Phys. C: Solid State Phys}, 19:3805, 1986.    

\bibitem{FouPou2010}
P.~Foury-Leylekian, J.~P. Pouget, Y.~J. Lee, R.~M. Nieminen, P.~Ordej\'{o}n, and E.~Canadell.
\newblock {\em Phys. Rev. B} 82:134116, 2010. 
  
\bibitem{LeeNie2003}
Y.~J. Lee, R.~M. Nieminen, P.~Ordej\'{o}n, and E.~Canadell.
\newblock {\em Phys. Rev. B} 67:180505(R), 2003. 

\bibitem{AlBuWh1985}
T.~A. Albright, J.~K. Burdett and M.-H.~Whangbo.
\newblock {\em Orbital Interactions in Chemistry}, 
\newblock John Wiley and Sons editors, New York, 1985.  
  
\bibitem{Mackenzie96}
A.~P. MacKenzie, S.~R. Julian, D.~C. Sinclair, and C.~T. Lin.
\newblock {\em Phys. Rev. B}, 53:5848, 1996.

\bibitem{Proust02}
C.~Proust, E.~Boaknin, R.~W.~Hill, L.~Taillefer, and A.~P.~Mackenzie.
\newblock {\em Phys. Rev. Lett.}, 89:147003, 2002.

\bibitem{Cooper09}
R.~A. Cooper, Y.~Wang, B.~Vignolle, O.~J. Lipscombe, S.~M. Hayden, Y.~Tanabe, T.~Adachi, Y.~Koike, M.~Nohara, H.~Takagi, C.~Proust, and N.~E. Hussey.
\newblock {\em Science}, 323:603, 2009.

\bibitem{Yonezawa08}
S.~Yonezawa, Y.~Maeno, P.~Auban-Senzier, C.~Pasquier, K.~Bechgaard, and
  D.~J\'erome.
\newblock {\em Phys. Rev. Lett.}, 100:117002, 2008.

\bibitem{Bourbonnais09}
C.~Bourbonnais and A.~Sedeki.
\newblock {\em Phys. Rev. B}, 80:085105, 2009.
\newblock arXiv.org:0904.2858.

\bibitem{Abanov03}
Ar.~Abanov, A.~V. Chubukov and J.~Schmalian.
\newblock {\em Adv. Physics}, 52:119, 2003.

\bibitem{Creuzet87b}
F.~Creuzet, C.Bourbonnais, L.G. Caron, D.~J\'erome, and A.~Moradpour.
\newblock {\em Synthetic Metals}, 19:277, 1987.

\bibitem{Brown08}
S.~E. Brown, P.~M. Chaikin and M.~J. Naughton.
\newblock In A.~G. Lebed, editor, {\em The Physics of Organic Superconductors and Conductors}, page~49-88, Springer Series in Materials Sciences, Springer Berlin, 2008.

\bibitem{Taillefer10}
L.~Taillefer.
\newblock {\em Annual Review of Condensed Matter Physics}, 1:51, 2010.

\end{thebibliography}
\end{document}